\documentclass[11pt]{article}

\usepackage{amsmath,amsfonts,amssymb,amsthm,graphicx,cite}

\newcommand{\ee}[1]{{\rm e}^{#1}}
\newcommand{\ii}{{\rm i}}
\newcommand{\dd}{{\rm d}}

\newcommand{\R}{{\mathbb R}}
\newcommand{\C}{{\mathbb C}}

\newcommand{\nonu}{\nonumber \\ \nopagebreak}
\newcommand{\bma}[1]{\begin{array}{#1}}
\newcommand{\ema}{\end{array}}

\newcounter{saveeqn}
\newcounter{App} 
\newcommand{\app}{%
\stepcounter{App}%
\setcounter{saveeqn}{\value{equation}}%
\setcounter{equation}{0}%
\renewcommand{\theequation}{\Alph{App}\arabic{equation}} }
\newcommand{\appende}{%
\setcounter{equation}{\value{saveeqn}}%
\renewcommand{\theequation}{\arabic{equation}}  }
\newcommand{\alpheqn}{%
\stepcounter{equation}
\setcounter{saveeqn}{\value{equation}}%
\setcounter{equation}{0}%
\renewcommand{\theequation}{\arabic{saveeqn}\alph{equation}} }
\newcommand{\reseteqn}{\setcounter{equation}{\value{saveeqn}\noindent}%
\renewcommand{\theequation}{\arabic{equation}} }
\newcounter{asaveeqn}

\begin{document}
\begin{flushright}
\vspace{.4cm}
May 12, 2004
\end{flushright}
\vspace{.4cm}
 
\begin{center}
 
{\Large\bf Exact solutions of two complementary 1D quantum
many-body systems on the half-line}

\vspace{1 cm}
 
{\large Martin Halln\"as and Edwin Langmann}\\
\vspace{0.3 cm}
{\em Mathematical Physics, Department of
Physics, KTH, AlbaNova, SE-106 91 Stockholm, Sweden}\\
\end{center}

\vspace{0.3cm}

\pagenumbering{arabic}

\begin{abstract}
We consider two particular 1D quantum many-body systems with local
interactions related to the root system $C_N$.  Both models describe
identical particles moving on the half-line with non-trivial boundary
conditions at the origin, and they are in many ways complementary to
each other. We discuss the Bethe Ansatz solution for the first model
where the interaction potentials are delta-functions, and we find that
this provides an exact solution not only in the boson case but even
for the generalized model where the particles are distinguishable.  In
the second model the particles have particular momentum dependent
interactions, and we find that it is non-trivial and exactly solvable
by Bethe Ansatz only in case the particles are fermions. This latter
model has a natural physical interpretation as the non-relativistic
limit of the massive Thirring model on the half-line.  We establish a
duality relation between the bosonic delta-interaction model and the
fermionic model with local momentum dependent interactions. We also
elaborate on the physical interpretation of these models. In our
discussion the Yang-Baxter relations and the Reflection equation play
a central role.
\end{abstract}

\section{Introduction}
Quantum mechanical models with interactions are, in general, very
difficult to solve, but there exist a few important cases where exact
solutions are available, allowing them to be understood completely.  A
prominent example is the delta interaction in one dimension which, in
the simplest two-particle case, is defined by the Hamiltonian
\begin{equation}
\label{h1}
	H=-\partial_x^2 + c\, \delta(x)
\end{equation}
where $c$ is a real coupling constant and $x\in\R$ the relative
coordinate of the two particles, $x=x_1-x_2$.  This latter model is
popular because it allows for an explicit solution by simple means:
since the delta interaction is restricted to $x=0$, it only manifests
itself in the non-trivial boundary conditions for eigenfunctions
$\psi(x)$ of $H$,
\begin{eqnarray}
\label{bc1} 
	\psi(0^+) = \psi(-0^+)\nonu
	\psi'(0^+)-\psi'(-0^+) = c \, \psi(0^+) , 
\end{eqnarray}
and these can be easily accounted for (we write $\psi(\pm 0^+)$ short
for the left-and right limits $\lim_{x\downarrow 0}\psi(\pm x)$, and
similarly for the derivative $\psi'$).  The natural generalization of
this model to an arbitrary number $N$ of identical particles defines a
prominent exactly solvable quantum many-body system which, in the
boson case, was solved by Lieb and Liniger \cite{LL} and, for the
general case of distinguishable particles, by Yang \cite{Y} in
a seminal paper where the Yang-Baxter relations first appeared.

Interactions localized at points have been studied extensively using
the mathematical theory of defect indices; see \cite{AGHH} and
references therein. From these studies it is well-known that the delta
interaction is only one of many possible local interactions, and a
general such interaction can be characterized by four real coupling
parameters. This can be easily understood as follows: for a 1D
Hamiltonian $H=-\partial_x^2+\hat v$ with an interaction $\hat v$
localized at $x=0$ all eigenfunctions $\psi(x)$ should be smooth
everywhere except at $x=0$, and $(H\psi)(x) = -\psi''(x)$ for non-zero
$x$. Requiring $H$ to be self-adjoint leads to the following
condition,
\begin{equation} 
	\int_{|x|>0}\dd x\, \Bigl( \overline{\phi''(x)} \psi(x)
	-\overline{\phi(x)} \psi''(x) \Bigr) = 0
\end{equation}
for arbitrary wave functions $\phi$ and $\psi$, or equivalently
\begin{equation} 
\label{cond}
	[\overline{\phi'}\psi - \overline{\phi} \psi']_{x=0^+} =
	[\overline{\phi'}\psi - \overline{\phi} \psi']_{x=-0^+} .
\end{equation}
General boundary conditions are of the form
\begin{eqnarray}
	\psi(0^+) &=& u_{11} \psi(-0^+) + u_{12} \psi'(-0^+)\nonu
	\psi(0^+) &=& u_{21} \psi(-0^+) + u_{22} \psi'(-0^+)
\end{eqnarray}
(and similarly for $\phi$, of course) and are thus parameterized by
four complex parameters $u_{jk}$ which, when imposing (\ref{cond}),
are reduced to two complex, or equivalently, four real parameters.
The boundary conditions in Eq.\ (\ref{bc1}) are obviously contained in
this class of boundary conditions, but there are others, most
prominently
\begin{eqnarray}
\label{bc2} 
	\psi'(0^+) = \psi'(-0^+)\nonu \psi(0^+) -
	\psi(-0^+) = \lambda \psi'(0^+)
\end{eqnarray}
which often has been referred to as delta-prime interaction; see e.g.\
Section I.4 in \cite{AGHH}. Recently it was shown that these latter
boundary conditions arise naturally from the Hamiltonian
\begin{equation}
\label{h2}
	H=-\partial_x^2 + \lambda\partial_x\delta(x)\partial_x 
\end{equation}
where the second term has a physical interpretation as a local
interaction depending also on the momentum $\hat p=-\ii \partial_x$
\cite{GLP}. The $N$-body generalization of this model is exactly
solvable by Bethe Ansatz in the indistinguishable particle case when
the particles are either bosons or fermions \cite{GLP,CS} but,
different from the delta interaction case, not in the general case of
distinguishable particles \cite{GLP}.  Still, this model is
complementary to the model with the delta interactions for at least
three different reasons \cite{GLP}: firstly, for indistinguishable
particles, the delta interaction model is known to be interesting only
for bosons (since the delta interaction is trivial on fermion wave
functions), whereas the $\hat p\delta\hat p$-interaction is trivial
for bosons and non-trivial for fermions. Secondly, while the delta
interaction model for bosons can be obtained as the non-relativistic
limit of the quantum sine Gordon model, the $\hat p\delta\hat
p$-interaction model for fermions naturally arises as the
non-relativistic limit of the massive Thirring model. Thirdly, there
exists an interesting weak coupling duality between the fermionic
$\hat p\delta\hat p$-interaction model and the bosonic
delta-interaction model.

As is well-know, exactly solvable many-body systems of particles
moving on the full real line are naturally associated with the root
system $A_{N-1}$, and they often allow for extensions to other root
systems such that the exact solubility is preserved \cite{OP}.  An
early example was given by Gaudin who solved the $C_N$ root system
variant of the delta interaction model for bosons \cite{Ga}, while the
general case of this model for arbitrary root systems and
distinguishable particles was treated by Sutherland \cite{Su}.  As
pointed out by Cherednik \cite{Ch}, models related to the root system
$C_N$ describe interacting particles on the half line, and the exact
solubility requires the so-called Reflection equation to be added to
the Yang-Baxter relations. The Reflection equation has played a
central role in many exactly solvable systems with a boundary; see
e.g.\ \cite{Sk} and the review \cite{K}.

In this paper we consider the Bethe Ansatz solution of the $C_N$
versions of the models discussed above. Similarly as for the $A_{N-1}$
case we find that the delta interaction model is exactly solvable in
this way even for distinguishable particles, but for the model with
momentum independent interactions we obtain its exact solution only
for indistinguishable particles. We also elaborate on the physical
interpretation of these models as describing particles on the
half-line with non-trivial boundary conditions at the origin.

To be more specific, the models we discuss in the paper are defined by
the following Hamiltonians,
\begin{equation}
\label{hamiltonianCN}
	H = -\sum_{j=1}^N\partial_{x_j}^2 + 2c_1\sum_{j<k}\lbrack\delta(x_j -
	x_k) + \delta(x_j + x_k)\rbrack + c_2\sum_{j=1}^N\delta(x_j)
\end{equation}
(delta interactions) and
\begin{eqnarray}
\label{hamiltonianCN2} 
	H & = & -\sum_{j=1}^N\partial_{x_j}^2 +
	2\lambda_1\sum_{j<k}\lbrack (\partial_{x_j} -
	\partial_{x_k})\delta(x_j - x_k)(\partial_{x_j} -
	\partial_{x_k}) +\nonumber\\ && + (\partial_{x_j} +
	\partial_{x_k})\delta(x_j + x_k)(\partial_{x_j} +
	\partial_{x_k})\rbrack +
	4\lambda_2\sum_{j=1}^N\partial_{x_j}\delta(x_j)\partial_{x_j}
\end{eqnarray}
(local momentum dependent interactions). For simplicity we assume all
coupling constants positive so that there are no bound states.
Mathematically, the model in Eq.\ (\ref{hamiltonianCN}) is the $C_N$
variant of the model solved by Yang \cite{Y}, and Eq.\
(\ref{hamiltonianCN2}) defines the $C_N$ variant of the model
discussed in \cite{GLP}.

The plan of the rest of this paper is as follows.  In Section~2 we
consider the $C_N$ delta-interaction model, starting by deriving the
boundary conditions and thus turning the Schr\"odinger equation
$H\psi=E\psi$ into a well-defined mathematical problem. We proceed to
the Bethe Ansatz solution of this model where the Yang-Baxter
relations and the Reflection equation play a central role. We conclude
the section by elaborating on the physical interpretation of this
model. In Section~3 we discuss the $C_N$ variant of the $p\delta
p$-interaction model, in large parts paralleling our discussion for
the delta-interaction in Section~2. We also present a duality relation
between the fermionic $p\delta p$-interaction model and the bosonic
delta-interaction model. Appendix A gives some details on the
verification of the Yang-Baxter relations and the Reflection equation.
Appendix B contains a few mathematical facts about the Weyl group of
$C_N$, and Appendix C gives some details on the physical
interpretation of these models.

\section{Delta-interaction}
In this section we provide the exact solution of the $C_N$
delta-interaction in the case of distinguishable particles and
elaborate on its physical interpretation.

\subsection{Boundary conditions}
The Hamiltonian (\ref{hamiltonianCN}) defining the $C_N$
delta-interaction model is only formal, and to determine its
eigenfunctions we must first convert it into a set of boundary
conditions.

For completeness we start by discussing the Hamiltonian $H$ in Eq.\
(\ref{h1}), which can be regarded also as the one-particle case of the
Hamiltonian in Eq.\ (\ref{hamiltonianCN}), $N=1$. The first step to
find the eigenfunction $\psi$ of $H$ is to note that the equation
$H\psi=E\psi$ for all $x$ is equivalent to $-\psi''=E\psi$ for $x\neq
0$ together with the boundary conditions in Eq.\ (\ref{bc1}). These
boundary conditions are obtained by integrating the equation
$H\psi=E\psi$ twice: first from $x=-0^+$ to $x>0$ and then once more
from $x=-0^+$ to $x=0^+$ yields the first condition in Eq.\
(\ref{bc1}), and integrating from $x=-0^+$ to $x=0^+$ yields the
second condition in Eq.\ (\ref{bc1}). Thus in this case there are two
regions free of interactions, $x<0$ and $x>0$, linked to each other by
the boundary condition at $x=0$.

For general $N$, the interaction-terms of the Hamiltonian $H$ in Eq.\
(\ref{hamiltonianCN}) are restricted to $x_j=\pm x_k$ and $x_j=0$ for
$1\leq j<k\leq N$, and the eigenfunctions $\psi$ of $H$ therefore obey
the simple equation
\begin{equation} 
\label{freeEq} 
	\left(\sum_{j=1}^N \partial_{x_j}^2 + E\right)\psi(x_1,\ldots,x_N) = 0
	\quad \mbox{ for $x_j \neq \pm x_k$ and $x_j\neq 0$,} 
\end{equation}
and for each of the boundaries of the interaction free regions one
gets a pair of boundary conditions similarly to the ones for $N=1$,
\alpheqn
\begin{eqnarray}
\label{boundCond1}
	\psi\arrowvert_{x_j = \pm x_k + 0^+} &=& \psi\arrowvert_{x_j =
\pm x_k - 0^+} \nonu (\partial_{x_j} -
\partial_{x_k})\psi\arrowvert_{x_j = \pm x_k + 0^+} - (\partial_{x_j}
- \partial_{x_k})\psi\arrowvert_{x_j = \pm x_k - 0^+} &=&
2c_1\psi\arrowvert_{x_j = \pm x_k -0^+} \\ \label{boundCond5}
\psi\arrowvert_{x_j = + 0^+} &=& \psi\arrowvert_{x_j = -0^+} \nonu
\partial_{x_j}\psi\arrowvert_{x_j = 0^+} -
\partial_{x_j}\psi\arrowvert_{x_j = -0^+} &=& c_2\psi\arrowvert_{x_j
=0^+}
\end{eqnarray}
\reseteqn \noindent (these conditions are obtained by a
straightforward generalization of the $N=1$ argument above, using
$\partial_{x_j} \pm \partial_{x_k} = 2\partial_{x_j\pm x_k}$).

Obviously there are now many more regions free of interactions.  One
such region is $0<x_1<x_2<\ldots < x_N ,$ and all others are obtained
from this by permuting the particle labels, $j\to pj$ with $p\in S_N$
(= permutation group), and/or reflecting some of the coordinates,
$x_j\to -x_j$. Thus all regions free of interactions can be
characterized as follows,
\begin{equation}
	0< \sigma_1 x_{p1} < \sigma_2 x_{p2} < \ldots < \sigma_N
	x_{pN} <\infty
\end{equation}
where $\sigma_j=\pm 1$ and $p\in S_N$; we will refer to these regions
as {\em wedges}. It is important to note that they can be labeled by
elements $Q$ in the group
\begin{equation} 
\label{WN} 
	W_N \, := \, (\mathbb{Z}/2\mathbb{Z})^N\rtimes S_N 
\end{equation}
where the first factor corresponds to the reflections while the second
factor corresponds to the permutations of the coordinates,
\begin{equation} 
\label{Qx} 
	x_{Qj} = \sigma_j x_{pj}\quad \mbox{ for
	$Q=(\sigma_1,\ldots,\sigma_N;p)\in W_N$ with $\sigma_j\in\{ \pm 1\}$
	and $p\in S_N$.}  
\end{equation}
In the sequel we will therefore use the following convenient notation for
the wedges,
\begin{equation}
\label{wedge} 
	\Delta_Q:\quad 0<x_{Q1}<x_{Q2}<\ldots <x_{QN}
\end{equation} 
with $Q\in W_N$. It is interesting to note that the group $W_N$ is
isomorphic to the Weyl group of the root system $C_N$; see e.g.\ 
\cite{OP1}.

\subsection{Bethe Ansatz}
Using the boundary conditions deduced in the previous section we now
proceed to determine all eigenfunctions of the $C_N$
delta-interaction, starting by recalling the physical motivation of
the Bethe Ansatz below.  For that we first consider the Hamiltonian
$H$ in Eq.\ (\ref{h1}). In this case there are eigenfunctions
$\psi(x)=\exp(\ii kx)$ for $x<0$ which are equal to a particular
linear combination of $\exp(\ii kx)$ and $\exp(-\ii kx)$ for
$x>0$. This can be interpreted as scattering by the delta interaction
$\propto \delta(x)$ where a plane wave is partly transmitted and
partly reflected.  Regarding $H$ in Eq.\ (\ref{h1}) as a two particle
Hamiltonian with $x=x_1-x_2$ the relative coordinate and
$k=(k_1-k_2)/2$ the relative momentum, we can interpret this very fact
as scattering of a plane wave solution $\exp(\ii k_1 x_1 + \ii k_2
x_2)$ into a linear combination of this wave and another one where the
particle momenta $k_1$ and $k_2$ are exchanged, $\exp(\ii k_2 x_1 +
\ii k_1 x_2)$. This suggests that an eigenfunction $\psi$ of the
$N$-particle Hamiltonian in Eq.\ (\ref{hamiltonianCN}) which is equal
to a plane wave $\exp(\ii \sum_{j=1}^N k_j x_j)$ in one wedge
$\Delta_Q$ (\ref{wedge}) will be transformed into a linear combination
of plane waves $\exp(\ii \sum_{j=0}^N \tilde k_j x_j )$ in any other
wedge where $\tilde k_j = \sigma_j k_{pj}$, with $\sigma_j=\pm 1$
resulting from the interactions $\propto \delta(x_j)$ which can invert
momenta, $k_j\to -k_j$, and $p\in S_N$ resulting from the interactions
$\propto \delta(x_j - x_\ell)$ which can interchange momenta, $k_j
\leftrightarrow k_\ell$.

We thus see that the group in Eq.\ (\ref{WN}) naturally appears again,
$\tilde k_j = k_{Pj}$ for some $P\in W_N$, and the discussion above
suggests the following Bethe Ansatz for the eigenfunctions of the
Hamiltonian $H$ in Eq.\ (\ref{hamiltonianCN}),
\begin{eqnarray}
\label{betheAnsatz}
	\psi(x) = \sum_{P\in W_N} A_P(Q)\,\ee{\ii k_P\cdot x_Q}\quad 
	\mbox{ for $0<x_{Q1}<x_{Q2} < \ldots < x_{QN}$}	
\end{eqnarray} 
with $x=(x_1,\ldots,x_N)$ and $k_P\cdot x_Q\equiv \sum_{j=1}^N k_{Pj}
x_{Qj}$, for all $Q\in W_N$. The corresponding eigenvalue is obviously
$E=\sum_{j=1}^N k_j^2$.

One now has to take into account the boundary conditions in
(\ref{boundCond1},b). For each $Q\in W_N$, the wedge $\Delta_Q$ 
(\ref{wedge}) participates in $N$ boundaries: $x_{Qi} = x_{Q(i+1)}$ for
$i=1,2,\ldots (N-1)$ and $x_{Q1}=0$, and for each of these boundaries
we will get two conditions. More specifically, the boundary at
$x_{Qi} = x_{Q(i+1)}$ is between the wedges $\Delta_Q$ and
$\Delta_{QT_i}$ where $T_i\in W_N$ is the transposition interchanging
$i$ and $(i+1)$, and the conditions implied by Eq.\ (\ref{boundCond1})
for $j=Qi$ and $k=Q(i+1)$ are \alpheqn
\begin{eqnarray}
\label{cc1a}
	A_P(Q) + A_{PT_i}(Q) = A_{P}(QT_i) + A_{PT_i}(QT_i) \nonu
	\ii(k_{Pi} -k_{P(i+1)} ) [A_{PT_i}(QT_i) - A_P(QT_i) +
	A_{PT_i}(Q) -A_P(Q) ] = 2 c_1 [ A_{P}(Q) + A_{PT_i}(Q)] .
\end{eqnarray}
The boundary at $x_{Q1}=0$ is between the wedges $\Delta_{Q}$ and
$\Delta_{QR_1}$ with $R_1\in W_N$ the reflection of the first
argument, i.e., $x_{R_1 j} = x_j$ for $j\neq 1$ and $-x_j$ for $j=1$,
and the conditions at $x_{Q1}=0$ implied by Eq.\ (\ref{boundCond5}) for
$j=Q1$ are,
\begin{eqnarray}
\label{cc1b}  
	A_P(Q) + A_{PR_1}(Q) = A_{P}(QR_1) + A_{PR_1}(QR_1)\nonu
	\ii k_{P1}[A_P(Q) - A_{PR_1}(Q) + A_P(QR_1) - A_{PR_1}(QR_1)] =
	c_2\lbrack A_P(QR_1) + A_{PR_1}(QR_1)\rbrack .	
\end{eqnarray}
\reseteqn
We thus have $2N(2^N N!)^2$ linear, homogeneous equations for the
$(2^N N!)^2$ coefficients $A_P(Q)$. The following beautiful argument
due to Yang \cite{Y} shows that this system of equations has
enough non-trivial solutions and, at the same time, gives a recipe to
compute all the $A_P(Q)$.

For that it is important to note that $W_N$ plays a third role:
defining
\begin{equation}
  (\hat R)_{Q, Q^{\prime}} = \delta_{Q^{\prime},QR}
\end{equation}
one can write
\begin{equation}
\label{transPropCN}
  A_P(QR) = \sum_{Q^{\prime}\in W_N}(\hat R)_{Q,
  Q^{\prime}}A_P(Q^{\prime}) = (\hat RA_P)(Q)
\end{equation}
where the first equality is a trivial consequence of the definition,
and in the second we interpret $(\hat R)_{Q, Q^{\prime}}$ as elements
of an $n\times n$ matrix $\hat R$ with $n=2^NN!$ the rank of $W_N$.
These matrices obviously define a representation $R\to \hat R$ of
$W_N$ acting on the coefficients $A_P(Q)$. It is worth noting that
this is identical with the so called (right) regular representation of
$W_N$.

We can therefore insert $A_{PT_i}(QT_i) = (\hat T_i A_{PT_i})(Q)$ in
Eq.\ (\ref{cc1a}), and by a simple computation show that these latter
equations are equivalent to
\begin{equation}
\label{yOpRel}
	A_P=Y_i(k_{P(i+1)}-k_{Pi})A_{PT_i}
\end{equation}
where we have introduced the operator
\begin{equation}
\label{Yi1}
	Y_i(u)=\frac{\ii u\hat{T}_i + c_1\hat{I}}{\ii u - c_1} 
\end{equation}
and interpret $A_P$ as a vector with $2^NN!$ elements $A_P(Q)$.  In
the same way we can rewrite the conditions in Eq.\ (\ref{cc1b}) using
$A_{PR_1}(QR_1) = (\hat{R}_1A_{PR_1})(Q)$,
\begin{equation}
\label{zOpRel}
	A_P = Z(2k_{P1})A_{PR_1}
\end{equation}
with the operator
\begin{equation}
\label{Z1} 
	Z(u) = \frac{\ii u\hat{R}_1 + c_2\hat{I}}{\ii u - c_2} . 
\end{equation}

It is well-known that the group $W_N$ is generated by the reflection
$R_1$ and the transpositions $T_i$ (see e.g.\ page 21 in
\cite{GP}). Thus one can use the identities in Eqs.\ (\ref{yOpRel}),
(\ref{zOpRel}) and (\ref{transPropCN}) to calculate recursively all
coefficients $A_P(Q)$ from $A_I(I)$ using the operators $Z$ and $Y_i$
above.  It is important to note that there is a possible inconsistency
arising from the fact that the representation of an element $P$ in
$W_N$ as a product of the $T_i$'s and $R_1$ is not unique. However,
any two such representations can be converted into each other by using
the defining relations of the group $W_N$, \alpheqn
\begin{eqnarray}
	T_iT_i = 1,\qquad T_iT_j = T_jT_i,\qquad \textrm{for} \
	|i-j|>1\nonu T_iT_{i+1}T_i = T_{i+1}T_iT_{i+1}\\ R_1R_1 =
	1,\qquad R_1T_i = T_iR_1,\qquad \textrm{for} \ i>1\nonu 
	R_1T_1R_1T_1 = T_1R_1T_1R_1.
\end{eqnarray}
\reseteqn Thus no inconsistency can arise provided that \alpheqn
\begin{eqnarray}
\label{con1} 
	A_{PT_iT_i}(Q) = A_P(Q),\qquad A_{PT_iT_j}(Q) =
	A_{PT_jT_i}(Q),\qquad \textrm{for} \ |i-j|>1\nonu 
	A_{PT_iT_{i+1}T_i}(Q) = A_{PT_{i+1}T_iT_{i+1}}(Q)\\
	A_{PR_1R_1}(Q) = A_P(Q),\qquad A_{PR_1T_i}(Q) =
	A_{PT_iR_1}(Q),\qquad \textrm{for} \ i>1\nonu 
	A_{PR_1T_1R_1T_1}(Q) = A_{PT_1R_1T_1R_1}(Q)
\end{eqnarray}
\reseteqn for all $P, Q\in W_N$. Using the recurrence relations
(\ref{yOpRel}) and (\ref{zOpRel}) one finds that these conditions hold
true if and only if the following operator relations are fulfilled,
\alpheqn
\begin{eqnarray}
	Y_i(-u)Y_i(u) = I,\qquad Y_i(u)Y_j(v) = Y_j(v)Y_i(u),\qquad
	\textrm{for} \ |i - j|>1\nonu 
\label{yangBaxter}
	Y_i(v)Y_{i+1}(u + v)Y_i(u) =
	Y_{i+1}(u)Y_i(u + v)Y_{i+1}(v)\\  Z(-u)Z(u) =
	I,\qquad Z(u)Y_i(v) = Y_i(v)Z(u),\qquad \textrm{for} \
	i>1\nonu 
\label{reflEq}
        Z(2v)Y_1(u + v)Z(2u)Y_1(u - v) = Y_1(u -
	v)Z(2u)Y_1(u + v)Z(2v)
\end{eqnarray}
\reseteqn for all real $u$ and $v$. The validity of this system of
equations is necessary and sufficient in order for the Bethe Ansatz
above to be consistent and the model at hand to be exactly solvable.
The first three relations are the so called {\em Yang-Baxter
relations}, and the last one is the {\em Reflection equation}.  The
validity of these relations for arbitrary $\hat T_i$ and $\hat R_1$
can be checked by straightforward but somewhat tedious computations
(of course, the validity of the Yang-Baxter relation in this case is
known since a long time \cite{Y}, and this seems to be the case also
for the Reflection equation \cite{Su,Sk}, but for completeness we
provide the essential steps in the verification in Appendix A.1).

Thus the Bethe Ansatz (\ref{betheAnsatz}) is consistent even in the
general case of distinguishable particles, and we can calculate all
coefficients $A_P$ from $A_I$ using the recurrence relation
\begin{equation}
\label{recurrenceRelCN}
	A_P = \mathcal{W}_P(k)A_I
\end{equation}
where $\mathcal{W}_P(k)$ is a product of the operators $Y_i(k_{P(i+1)}
- k_{Pi})$ and $Z(2k_{P1})$ obtained by using repeatedly
(\ref{yOpRel}) and (\ref{zOpRel}).

Interesting special cases of this solution are when the particles are
indistinguishable, i.e., when the particles are fermions of bosons. In
the former case $\hat T_i=-I$, and Eq.\ (\ref{Yi1}) implies $Y_i(u)=-I$
independent of the coupling constant $c_1$. This shows that the delta
interaction is trivial for fermions. In the boson case we have $\hat
T_i=+I$, and $Y_i(u)$ is a non-trivial phase. As discussed in more
detail below, there are two different boson cases with different
physical interpretations, namely $\hat R_1= -I$ and $\hat R_1=+I$.

\subsection{Physical interpretation}
As is well-known, the $C_N$ delta-interaction model describe
interacting particles on the half-line with particular boundary
conditions at the origin \cite{Ch}. However, the general solution of
the $C_N$ delta-interaction model without any restrictions includes
many more eigenfunctions than any model on the half line, and the
relation between these models is therefore not completely obvious. In
this section we discuss the relation of these models in more
detail. We also give a physical interpretation of the boundary
conditions which occur as limits of particular external potentials
restricting the particles to the half line.

As discussed in Appendix B, in any irrep of the group $W_N$ the
reflections $R_j$ of the particle coordinate $x_j$ are represented
either by $\hat R_j=+1$ or $-1$.  For simplicity we now discuss in
more detail the cases where all $\hat R_j$ are the same, either $+1$
or $-1$, which from a physical point of view are the most interesting
cases. As discussed in Appendix B, these irreps of $W_N$ can be rather
easily understood since they are related in a simple way to irreps of
$S_N$.  Thus we can impose the following restriction on the
eigenfunctions $\psi$ of the Hamiltonian in Eq.\
(\ref{hamiltonianCN}),
\begin{equation}
\label{Rpm1}  
	(\hat R_j\psi)(x_1,\ldots,x_j,\ldots,x_N) \equiv 
	\psi(x_1,\ldots,-x_j,\ldots,x_N) = \pm
	\psi(x_1,\ldots,x_j,\ldots,x_N) . 
\end{equation}
With that assumption we can restrict ourselves to $x_j>0$, and the
boundary conditions in Eq.\ (\ref{boundCond1}) and 
Eq.\ (\ref{boundCond5}) reduce to 
\alpheqn
\begin{eqnarray}
	\psi\arrowvert_{x_j = x_k + 0^+} &=& \psi\arrowvert_{x_j = x_k -
	0^+}\nonu (\partial_{x_j} - \partial_{x_k})\psi\arrowvert_{x_j =
	x_k + 0^+} - (\partial_{x_j} -
	\partial_{x_k})\psi\arrowvert_{x_j = x_k - 0^+} &=&
	2c_1\psi\arrowvert_{x_j = x_k + 0^+} , 
\end{eqnarray}
and 
\begin{eqnarray}
\label{BCorigin} 
	\bma{rll} 2\partial_{x_j}\psi\arrowvert_{x_j = 0^+} =&
	c_2\psi\arrowvert_{x_j = 0^+} & \mbox{ for $\hat R_j=+1$} \\
	\psi\arrowvert_{x_j = 0^+} =& 0 \quad & \mbox{ for $\hat
	R_j=-1$} \ema ,
\end{eqnarray}
\reseteqn respectively. These are exactly the boundary conditions
obtained from the Hamiltonian
\begin{equation} 
	H_{0} = -\sum_{j=1}^N\partial_{x_j}^2 + 2c_1\sum_{j<k}
	\delta(x_j - x_k) \label{Hhalf} 
\end{equation}
describing particles on the half-line, $x_j>0$, and the boundary
conditions at the origin given in Eq.\ (\ref{BCorigin}).

It is also interesting to note that these later boundary conditions
are obtained by allowing the particles to move on the full line,
$x_j\in\R$, and adding a particular external potential $\sum_j V(x_j)$
to the Hamiltonian in Eq.\ (\ref{Hhalf}) which effectively constrains
the particles to the half line $x_j>0$. To be specific, these
potentials are given by
\begin{equation} 
	V(x) = \left\{ \bma{cc} V_0\Theta(-x) +(c_2/2
	-\sqrt{V_0})\delta(x) &\mbox{ if $\hat R_j=+1$ } \\ V_0\Theta(-x)
	&\mbox{ if $\hat R_j=-1$} \ema \right. ,
\end{equation}
where $\Theta(-x)$ is the Heaviside function (equal to one for $x<0$
and zero otherwise), and one has to take the strong coupling limit
$V_0\to\infty$: as shown in Appendix C, in this latter limit the
eigenfunctions of the Hamiltonian $H_0+ \sum_j V(x_j)$ on the full
line, $x_j\in \R$, coincide with the ones of $H_0$ on the half-line,
$x_j>0$, and the boundary conditions in Eq.\ (\ref{BCorigin}).

As already mentioned, the most important cases in applications are the
ones we have considered here, i.e., where all the $\hat R_j$ are the
same. Nevertheless it would be of interest to consider the
implications of allowing the $\hat R_j$ to take on different values,
in effect dividing the particles into two groups distinguished by
their interactions with the boundary.

\section{Local momentum-dependent interaction}
In this section we discuss the model with local momentum dependent
interactions defined by the Hamiltonian in Eq.\
(\ref{hamiltonianCN2}). While most of our discussion is in parallel
with the one for the delta interaction model in the previous section,
we find that the Bethe Ansatz is consistent only for the
indistinguishable particle case. We also present a duality relation
between fermionic variant of the model here and the bosonic model
discussed in the previous section.

\subsection{Boundary conditions}
We start by considering the $N=1$ Hamiltonian $H$ in Eq.\
(\ref{h2}). To obtain the corresponding boundary conditions we first
integrate from $x=-0^+$ to $x=0^+$ which yields the first condition in
Eq.\ (\ref{bc2}), and integrating from $x=-0^+$ to $x>0$ and then once
more from $x=-0^+$ to $x=0^+$ yields the second condition. As in the
delta interaction case, the eigenfunctions $\psi$ of $H$ are then
determined by these conditions together with the equation
$-\psi''=E\psi$ for $x\neq 0$.  We note that the wave functions
$\psi(x)$ on which $H$ in Eq.\ (\ref{h2}) is defined can be
discontinuous at $x=0$, and to make sense of the interactions we have
implicitly used a regularization which amounts to replacing $\psi'(0)$
by $[\psi'(0^+)+\psi'(-0^+)]/2$ (this is discussed in more detail in
\cite{GLP})

It is straightforward to generalize this argument to the $N$-particle
case. Similarly as in the delta interaction case one finds that the
eigenfunctions $\psi$ of the Hamiltonian in Eq.\
(\ref{hamiltonianCN2}) are determined by Eq.\ (\ref{freeEq}) together
with the boundary conditions \alpheqn
\begin{eqnarray}
	(\partial_{x_j} - \partial_{x_k})\psi\arrowvert_{x_j = \pm x_k
	+ 0^+} &=& (\partial_{x_j} -
	\partial_{x_k})\psi\arrowvert_{x_j = \pm x_k - 0^+}
\label{boundCond21} 
	\nonu \psi\arrowvert_{x_j = \pm x_k + 0^+} -
	\psi\arrowvert_{x_j = \pm x_k - 0^+} &=& 2\lambda_1(\partial_{x_j} -
	\partial_{x_k})\psi\arrowvert_{x_j = \pm x_k - 0^+} \\
	\partial_{x_j}\psi\arrowvert_{x_j = 0^+} &=&
	\partial_{x_j}\psi\arrowvert_{x_j = -0^+}
\label{boundCond25}
	\nonu \psi\arrowvert_{x_j = 0^+} - \psi\arrowvert_{x_j = -0^+}
	&=& 4\lambda_2\partial_{x_j}\psi\arrowvert_{x_j = 0^+}.
\end{eqnarray}
\reseteqn

\subsection{Bethe Ansatz}
We now discuss the Bethe Ansatz for the eigenfunctions of the
Hamiltonian $H$ defined in Eq.\ (\ref{hamiltonianCN2}). Obviously much
of what we said for the delta interaction case carries over
straightforwardly to the present case. Due to the different boundary
conditions in Eqs.\ (\ref{boundCond21},b) Eqs.\ (\ref{cc1a},b) are
changed to \alpheqn
\begin{eqnarray} 
\label{cc2a}
\ii (k_{Pi} - k_{P(i+1)}) [A_{PT_i}(QT_i) - A_P(QT_i)] =
\ii(k_{Pi} - k_{P(i+1)}) [ A_P(Q)- A_{PT_i}(Q)] \nonu A_{P}(QT_i) +
A_{PT_i}(QT_i) - A_P(Q) - A_{PT_i}(Q) = 2\lambda_1 \ii(k_{Pi} -
k_{P(i+1)}) [ A_P(Q) - A_{PT_i}(Q)] \\
\label{cc2b}
\ii k_{P1}[A_P(Q) - A_{PR_1}(Q)] = \ii k_{P1} [
A_{PR_1}(QR_1)-A_P(QR_1)]\nonu A_P(Q) + A_{PR_1}(Q) - A_{P}(QR_1) -
A_{PR_1}(QR_1) = 4\lambda_2 \ii k_{P1} [ A_P(QR_1)- A_{PR_1}(QR_1)]
 .
\end{eqnarray} 
\reseteqn
We now also use Eq.\ (\ref{transPropCN}) to convert these in to the recurrence
relations
\begin{equation}
\label{yOpRel2}
	A_P = Y_i(k_{P_{i+1}} - k_{P_i})A_{PT_i} , 
\end{equation}
and similarly 
\begin{equation}
\label{zOpRel2}
	A_P = Z(2k_{P_1})A_{PR_1}
\end{equation}
where now
\begin{equation}
\label{Yi2}
	Y_i(u) = \frac{\ii u\hat{I} - 1/\lambda_1\hat{T_i}}{\ii u -
	1/\lambda_1}
\end{equation}
and
\begin{equation}
\label{Z2} 
	Z(u) = \frac{\ii u\hat{I} - 1/\lambda_2\hat{R}_1}{\ii u - 1/\lambda_2}.
\end{equation}
As in the delta interaction case these relations allow to recursively
compute all coefficients $A_P$ in terms of $A_I$, and the conditions
for the absence of inconsistencies are identical to (\ref{con1},b) of
the delta-interaction case, leading to the Yang-Baxter relations
(\ref{yangBaxter}) and Reflection equation (\ref{reflEq}) but now with
the operators (\ref{Yi2}) and (\ref{Z2}). In contrast to the
delta-interaction case, we find that these consistency relations are
valid only if $\hat T_i = \pm I$ for all $i$ (see Appendix~A.2 for
details). We thus conclude that {\em the Bethe Ansatz is consistent
only if the particles are indistinguishable, i.e., $A_I$ is chosen
such that either $\hat T_i=I$ or $\hat T_i=-I$,} and in these two
cases we can compute all coefficients $A_P$ from $A_I$ as
\begin{equation}
\label{recurrenceRelCN2}
	A_P = \mathcal{W}_P(k)A_I
\end{equation} 
where $\mathcal{W}_P(k)$ is a product of operators $Y_i(k_{P_{i+1}} -
k_{P_i})$ and $Z(2k_{P_1})$ in Eqs.\ (\ref{Yi2}) and (\ref{Z2})
obtained by using repeatedly (\ref{yOpRel2}) and (\ref{zOpRel2}).

For $\hat T_i=+I$ we get from Eq.\ (\ref{Yi2}) that $Y_i(u)=I$
independent of $\lambda_1$, and we conclude that {\em the
momentum-dependent interaction is trivial for bosons}. However, for
$\hat T_i=-I$ (fermions) the $Y_i(u)$ are non­trivial phases. There
are two different fermions cases, namely $\hat R_1=\pm I$.

\subsection{Duality} 
It is interesting to note that there exists a simple duality relation
between the fermionic $\hat p\delta \hat p$ model considered here and
the bosonic $C_N$ delta-interaction model discussed in
Section~2. Since the operators $Y_i(u)$ and $Z(u)$ for the latter
model is identical with the ones of the fermions $\hat p\delta \hat p$
model upon the substitution $\lambda_1\rightarrow 1/c_1$ and
$\lambda_2\rightarrow 1/c_2$ (compare Eqs.\ (\ref{Yi1}) and (\ref{Z1})
for $\hat T_i=R_1=+I$ and Eqs.\ (\ref{Yi2}) and (\ref{Z2}) for $\hat
T_i=R_1=-1$), Eqs. (\ref{recurrenceRelCN}) and
(\ref{recurrenceRelCN2}) imply that
\begin{equation} 
	A_P^\delta\arrowvert_{\hat T_i=\hat R_1=+I} = A_P^{\hat p\delta \hat
	p}\arrowvert_{\hat T_i=\hat R_1= -I, \lambda_1\rightarrow
	1/c_1, \lambda_2\rightarrow 1/c_2} ,
\end{equation}
where $A_P^\delta$ are the coefficients of Section~2.2 and $ A_P^{\hat
p\delta \hat p}$ the ones in Section~3.2. This implies that {\em the
bosonic wave functions of the delta model in Section~2.2 and the
fermionic wave functions of the $\hat p\delta\hat p$-model in
Section~3.2 are identical when restricted to the fundamental wedge
\begin{equation}
	\Delta_I:\qquad 0<x_1<x_2<\ldots<x_N , 
\end{equation}
provided that the coupling constants of these models are related as
follows,
\begin{equation}
	\lambda_1=\frac1{c_1}\quad \mbox{ and } \quad \lambda_2=\frac1{c_2} .
\end{equation}
} This can be seen also more directly: assuming that the eigenfunction
$\psi$ of the Hamiltonian in Eq.\ (\ref{hamiltonianCN}) is bosonic,
$\hat T_i=\hat R=I$, it is enough to determine it in the fundamental
wedge. Moreover, the continuity conditions in Eqs.\
(\ref{boundCond1},b) are fulfilled automatically for boson wave
functions, whereas the conditions on the derivatives simplify to
\begin{eqnarray}
	(\partial_{x_j} - \partial_{x_{j+1}} - c_1)\psi\arrowvert_{x_j
	= x_k + 0^+} = 0 \nonu
\label{boundCondAN}
	(2\partial_{x_j} - c_2)\psi\arrowvert_{x_j = 0^+} = 0
\end{eqnarray}
for all $x$ in the fundamental wedge. In a similar manner one finds
that the fermionic eigenfunctions of the Hamiltonian in Eq.\
(\ref{hamiltonianCN2}), $\hat T_i=\hat R=-I$, are determined by the
very same conditions in Eq.\ (\ref{boundCondAN}) with $c_{1,2}$
replaced by $1/\lambda_{1,2}$.

This generalizes the duality previously observed in the $A_{N-1}$ case
\cite{GLP,CS} to the $C_N$ case.

\subsection{Physical interpretation}
As in the delta interaction case, one can restrict the eigenfunctions
$\psi$ of the Hamiltonian in Eq.\ (\ref{hamiltonianCN2}) by imposing
the conditions in Eq.\ (\ref{Rpm1}), reducing the boundary conditions
in Eqs.\ (\ref{boundCond21},b) to \alpheqn
\begin{eqnarray}
	(\partial_{x_j} - \partial_{x_k})\psi\arrowvert_{x_j = x_k +
	0^+} = (\partial_{x_j} - \partial_{x_k})\psi\arrowvert_{x_j =
	x_k - 0^+}\nonu \psi\arrowvert_{x_j = x_k + 0^+} -
	\psi\arrowvert_{x_j = x_k - 0^+} = 2\lambda_1(\partial_{x_j} -
	\partial_{x_k})\psi\arrowvert_{x_j = x_k + 0^+}
\end{eqnarray}
and 
\begin{eqnarray}
	\bma{rll} \partial_{x_j}\psi\arrowvert_{x_j = 0^+} =& 0 
	& \mbox{ for $\hat R_j=+1$} \\\psi\arrowvert_{x_j = 0^+} = & 
	2\lambda_2\partial_{x_j}\psi\arrowvert_{x_j = 0^+}
	& \mbox{ for $\hat R_j=-1$} \ema \label{BCorigin1} 
\end{eqnarray}
\reseteqn where $x_j>0$. This shows that the eigenfunctions of the $C_N$
Hamiltonian in Eq.\ (\ref{hamiltonianCN2}) with the restriction in Eq.\
(\ref{Rpm1}) are identical to the ones of the $A_{N-1}$ Hamiltonian
\begin{equation}
\label{Hhalf1}
	H_0  =  -\sum_{j=1}^N\partial_{x_j}^2 +
	2\lambda_1\sum_{j<k} (\partial_{x_j} -
	\partial_{x_k})\delta(x_j - x_k)(\partial_{x_j} -
	\partial_{x_k}) 
\end{equation}
restricted to the half-line, $x_j>0$, and the boundary conditions at the
origin given in Eq.\ (\ref{BCorigin1}).

Moreover, as shown in Appendix C.2, the eigenfunctions $\psi$ above
restricted to $x_j>0$ become identical to the ones of the Hamiltonian
$H_0 + \sum_j V(x_j)$ on the full real line, $x_j\in\R$, but with an
external potential
\begin{equation} 
	V(x) = \left\{ \bma{cc} V_0\Theta(-x)
        +\sqrt{V_0}\partial_x\delta(x) \partial_x &\mbox{ if $\hat
        R_j=+1$ } \\ V_0\Theta(-x) +
        2\lambda_2\partial_x\delta(x)\partial_x &\mbox{ if $\hat
        R_j=-1$} \ema \right.
\end{equation}
in the limit $V_0\to \infty$.

\section{Concluding remark} 
As discussed in the Introduction, there exists a 4-parameter family of
local interactions \cite{AGHH}, and the delta- and $\hat p\delta\hat
p$-interactions only correspond to one-parameter subfamilies each.  It
is therefore natural to ask: What about the other local interactions?
Are there other cases leading to exactly solvable models?  It is thus
interesting to note that there is a simple physical interpretation of
the four parameter family of local interactions which seems more
natural than the ones given before \cite{AGHH}: in the simplest case
they correspond to the following generalization of the Hamiltonians in
Eqs.\ (\ref{h1}) and (\ref{h2}),
\begin{equation} 
\label{H4} 
	H = -\partial_x^2 + c\delta(x) + \lambda \partial_x
	\delta(x)\partial_x + \gamma \partial_x \delta(x) -
	\overline{\gamma} \delta(x) \partial_x , 
\end{equation}
which obviously is the most general hermitian Hamiltonian with
interactions localized in $x=0$ and containing only derivatives up to
second order (higher derivatives than that do not lead to physically
acceptable boundary conditions). This Hamiltonian is formally
self-adjoint for arbitrary parameters $c,\lambda\in\R$ and
$\gamma\in\C$, and it indeed corresponds to the 4-parameter family of
local interactions mentioned above \cite{GHLP}. All these models have
natural generalizations to the many-body case, but there is only one
case where these latter models are known to be exactly solvable even
for distinguishable particles by the coordinate Bethe Ansatz:
$(c,\lambda,\gamma)=(c,0,0)$. It would be interesting to know if there
are other exactly solvable cases. We plan to come back to this
question elsewhere \cite{GHLP}. We only mention here that the
many-body generalization of the Hamiltonian in Eq.\ (\ref{H4})
describes identical particles only if $\gamma=0$, and to find exactly
solvable case for non-zero $\gamma$ therefore requires an extension of
Yang's method of solution \cite{Y} (which only works for identical
particle models).

\section*{Acknowledgments}
We would like to thank Harald Grosse, Jouko Mickelsson and Cornelius
Paufler for helpful discussions.  E.L. was supported in part by the
Swedish Science Research Council~(VR) and the G\"oran Gustafsson
Foundation.

\app
\section*{Appendix A. Verification of consistency relations}
In this appendix we sketch the verification of the consistency
relations in Eqs.\ (\ref{yangBaxter},b) (Yang-Baxter relations and the
Reflection equation).

\subsection*{A.1 Delta-interaction} 
We start by writing the operators $Y_i$ in the following way:
\begin{equation}
\label{Yii} 
	Y_i(u) = a(u) + b(u)\hat T_i
\end{equation}
where
\begin{equation}
\label{A1} 
	a(u) = \frac{c_1}{\ii u - c_1},\qquad b(u) = \frac{\ii u
	}{\ii u - c_1}.
\end{equation}
Inserting this expression into the equations in (\ref{yangBaxter})
results in a number of relations between the coefficients $a(u)$ and
$b(u)$, one for each equation and different permutation operator. Most
of them are trivially fulfilled, but the following ones are
non-trivial:
\begin{eqnarray}
\label{check1} 
	a(-u)a(u) + b(-u)b(u) = 1\nonu
	a(-u)b(u) + b(-u)a(u) = 0
\end{eqnarray} 
and 
\begin{eqnarray}
\label{check2} 
	b(v)a(u+v)a(u) + a(v)a(u+v)b(u) = a(u)b(u+v)a(v) . 
\end{eqnarray}
Inserting $a(u)$ and $b(u)$ from Eq.\ (\ref{A1}) they can be verified
by straightforward calculations. To verify Eq.\ (\ref{reflEq}) we
write the operator $Z$ as
\begin{equation}
	Z(u) = \tilde a(u) + \tilde b(u)\hat{R}_1 \label{ZZ} 
\end{equation}
where
\begin{equation}
	\tilde a(u) = \frac{c_2}{\ii u - c_2},\qquad \tilde b(u) =
	\frac{\ii u}{\ii u - c_2}.
\end{equation}
Substituting this and Eq.\ (\ref{Yii}) leads to the following
non-trivial relation,
\begin{eqnarray}
	\tilde b(2v)b(u + v)\tilde a(2u)a(u - v) + \tilde b(2v)a(u +
	v)\tilde a(2u)b(u - v) +\nonumber\\
	+\ \tilde a(2v)a(u + v)\tilde b(2u)b(u -
	v) = a(u - v)\tilde b(2u)b(u + v)\tilde a(2v)
\end{eqnarray}
in addition to
\begin{eqnarray}
	\tilde a(-u)\tilde a(u) + \tilde b(-u)\tilde b(u) = 1\nonu
	\tilde a(-u)\tilde b(u) + \tilde b(-u)\tilde a(u) = 0 , 
\end{eqnarray}
the validity of which follow from straightforward calculations.

We conclude that the Bethe Ansatz is consistent even for
distinguishable particles.

\subsection*{A.2 Local momentum-dependent interaction}
In this case we get $Y_i(u)$ as in Eq.\ (\ref{Yii}) but with
\begin{equation}
\label{A2} 
	a(u) = \frac{\ii u}{\ii u - 1/\lambda_1},\qquad b(u) =
	\frac{-1/\lambda_1 }{\ii u - \lambda_1}.
\end{equation}
With that the two equations in (\ref{check1}) hold true but the
equation in (\ref{check2}) does not. We therefore conclude that
{\em the Bethe Ansatz is not consistent for distinguishable
particles}.

For indistinguishable particles we have $\hat T_i=\pm I$ and the
Yang-Baxter relations in Eq.\ (\ref{yangBaxter}) are trivially
fulfilled. Moreover, in this case it is also easy to check that the
relations (\ref{reflEq}) hold true for $\hat R_1=\pm I$.

We conclude that the Bethe Ansatz is consistent in the
indistinguishable particle case but not in general.
 
\appende

\app
\section*{Appendix B. Representations of the group $W_N$}
In this appendix we discuss the irreducible representations of the
group $W_N\equiv (\mathbb{Z}/2\mathbb{Z})^N\rtimes S_N$. In particular
we will show the following. 

\bigskip

\noindent {\bf Fact:} {\em There exists a set of irreducible
representations of $W_N$ isomorphic to the irreducible representations
$\chi_{\pm}\otimes \rho$, where $\chi_{\pm}$ is a character
(irreducible representation) of the (normal) abelian subgroup
$(\mathbb{Z}/2\mathbb{Z})^N$ such that $\chi_{\pm}(R_j) = \pm 1$ for
all $j=1,2,\ldots,N$ (same sign for all $j$) and $\rho$ is an
arbitrary irreducible representation of the permutation group $S_N$}.

\bigskip

To show this we will use the notion of induced representations,
following Section~8.2 of \cite{Serre}.  We start by determining
the group of characters $X =
\textrm{Hom}((\mathbb{Z}/2\mathbb{Z})^N,\mathbb{C})$ of the subgroup
$(\mathbb{Z}/2\mathbb{Z})^N$. The fact that it is generated by the
reflections $R_j$ obeying the relations (see e.g.\ page 21 in \cite{GP})
\begin{equation}
	R_j^2 = I,\qquad j=1,2,\ldots,N
\end{equation}
implies that the characters $\chi\in X$ are functions such that
\begin{equation}
	\chi(R_j) = e^{in_j\pi},\qquad n_j\in\mathbb{Z}
\end{equation}
for all $j=1,2,\ldots,N$. The group $W_N$ acts on these characters by
\begin{equation}
	(w\chi)(R) = \chi(w^{-1}Rw),\qquad {\forall} w\in W_N, \chi\in
	X, R\in (\mathbb{Z}/2\mathbb{Z})^N.
\end{equation}
We now determine the orbits of the action of $S_N$ in $X$, represented
by a set $\chi_i$ where $i\in X/S_N$. Using the fact that the adjoint
action of $S_N$ permutes the reflections $R_j$, $T_{jk}R_jT_{jk} =
R_k$ with $T_{jk}$ the transposition interchanging $j$ and $k$, we
conclude that the orbits of $S_N$ in $X$ can be represented by the
characters
\begin{equation}
	\chi_k(R_j) =
	\left\{ \begin{array}{ll}
	1, & j>k\\
	-1, & j\leq k\end{array}\right.
\end{equation}
where $j,k=1,2,\ldots,N$. For each $i$ let $(S_N)_i$ be that subgroup
of $S_N$ consisting of all $P\in S_N$ such that $P\chi_i = \chi_i$,
and let further $\tilde W_i = (\mathbb{Z}/2\mathbb{Z})^N\cdot
(S_N)_i$. The structure of $\chi_i$ implies that $(S_N)_i = S_i\times
S_{N-i}$. The character $\chi_i$ can be extended to all of $\tilde
W_i$ by setting
\begin{equation}
	\chi_i(RP) = \chi(R),\qquad R\in (\mathbb{Z}/2\mathbb{Z})^N,
	P\in (S_N)_i.
\end{equation}
Now let $\rho_i$ be an irreducible representation of $(S_N)_i$ and
combine it with the canonical projection $\tilde W_i\rightarrow
(S_N)_i$ to yield an irreducible representation $\tilde{\rho}_i$ of
$\tilde W_i$. By taking the tensor product of $\chi_i$ and
$\tilde{\rho}_i$ we can now construct a set of irreducible
representations $\chi_i\otimes \rho_i$ of $\tilde W_i$. We denote the
corresponding induced representation of the whole of $W_N$ by
$\theta_{i,\rho_i}$. It follows from Proposition 25 in
 \cite{Serre} that all irreducible representations of $W_N$ are
isomorphic to such a representation $\theta_{i,\rho_i}$. In particular
setting $i=0$ and $i=N$ we arrive at the claim stated in the Fact at
the beginning of the section.  \appende

\app
\section*{Appendix C. Physical interpretation of boundary conditions}
In this appendix we substantiate the physical interpretation of the
boundary conditions of the $C_N$ models given in Sections 2 and 3 in
the main text.

\subsection*{C.1 Delta-interaction}
We first recall the eigenfunctions $\psi$ of the one particle
Hamiltonian in Eq.\ (\ref{h1}).  Since this Hamiltonian is invariant
under the reflection $x\rightarrow -x$ these eigenfunctions can be
chosen such that $\psi(x) = \pm\psi(-x)\equiv \psi_\pm(x)$, and they
can be computed using the Ansatz
\begin{equation}
	\psi_{\pm}(x) = \left\{ \bma{ll} \ee{-\ii kx} + A_\pm\ee{\ii kx} &
	\mbox{ for $x>0$ } \\ \pm \left( \ee{\ii kx} +
	A_\pm\ee{-\ii kx} \right) & \mbox{ for $x<0$ }\ema \right. , 
\label{scatteringAnsatz}
\end{equation}
and the boundary conditions in Eq.\ (\ref{bc1}) determine the
constants $A_\pm$ as follows,
\begin{equation}
	A_+ = \frac{\ii k + c/2}{\ii k - c/2},\qquad A_- = -1 
\end{equation}
with $A_-$ being independent of $c$ corresponding to the fact that the
delta interaction is trivial (i.e.\ invisible) for
fermions. Obviously, these eigenfunctions obey 
\begin{equation}
-\psi''_+(x) = k^2\psi_+(x) \quad \mbox{ for $x> 0 $ and } \quad
 \psi'(0^+) = (c/2) \psi(0^+)
\end{equation}
and 
\begin{equation}
-\psi''_-(x) = k^2\psi_-(x) \quad \mbox{ for $x> 0 $ and } \quad
 \psi_- (0^+) = 0 ,
\end{equation}
which is the simplest non-trivial case $N=1$ of the general
relation between the $C_N$ model and the $A_{N-1}$ model discussed in
Section~4.1.

We now show that these eigenfunctions $\psi_\pm(x)$ for $x>0$ are
identical to the ones of the Hamiltonians
\begin{equation}
	H_\pm = -\partial_x^2 + V_0\Theta(-x) + g_\pm \delta(x)  
\end{equation}
with
\begin{equation}
	g_+ = c/2 -\sqrt{V_0}\quad \mbox{ and } \quad g_- = 0  \label{gpm}  
\end{equation}
in the limit $V_0\to\infty$.  To show this we determine the
eigenfunctions $\phi_\pm$ of $H_\pm$ with the Ansatz
\begin{equation}
\label{scatteringAnsatz2}
	\phi_\pm =
		\left\{ \begin{array}{ll}
		\ee{-\ii kx} + B_\pm \ee{\ii kx}, & \mbox{ for $x>0$} \\
		C_\pm e^{\omega x}, & \mbox{ for $x<0$}
                \end{array}\right., 
\end{equation}
and by straightforward computations we find 
\begin{equation}
	B_\pm = \frac{\ii k + (\omega + g_\pm)}{\ii k - (\omega +
	g_\pm)}\quad \mbox{ and } \quad \omega = \sqrt{V_0 - k^2}
\end{equation}
for $V_0>k^2$. We thus see that
\begin{equation}
	A_\pm = \lim_{V_0\to\infty} B_\pm 
\end{equation}
provided that $g_\pm$ are chosen as in Eq.\ (\ref{gpm}). This shows
that the eigenfunctions $\phi_+$ of the Hamiltonian $H_+$ on the full
line in the limit $V_0\to\infty$ become equal to $\psi_+(x)$ for $x>0$
(and zero otherwise), and similarly for $\phi_-$, $\psi_-$ and $H_-$. 

This computation substantiates the physical interpretation of the
$C_N$ model in case $N=1$. However, since this interpretation only
involves the boundary conditions at $x_j=0$ which are not affected by
the inter-particle interactions, this argument immediately generalizes
to the $N>1$ particle case.

\subsection*{C.2 Local momentum dependent interaction}
The discussion for the Hamiltonian in Eq.\ (\ref{h2}) is completely
analogous to the one for the Hamiltonian in Eq.\ (\ref{h1}) given
above, and we therefore only write down the formulas which change.

Eq.\ (\ref{scatteringAnsatz}) determining the even and odd
eigenfunctions $\psi_\pm$ remains the same but $A_+$ and $A_-$ are
(essentially) interchanged,
\begin{equation}
	A_+ = 1 ,\quad A_- = \frac{\ii k + 1/2\lambda}{\ii k - 1/2\lambda} ,
\end{equation}
where now the boson eigenfunction is unaffected by the interaction.
Moreover, these eigenfunctions solve the following problems on the
half axis,
\begin{equation}
-\psi''_+(x) = k^2\psi_+(x) \quad \mbox{ for $x> 0 $ and } \quad
 \psi'_+(0^+) = 0
\end{equation}
and 
\begin{equation}
-\psi''_-(x) = k^2\psi_-(x) \quad \mbox{ for $x> 0 $ and } \quad
 \psi'_- (0^+) = 2\lambda \psi_-(0^+) . 
\end{equation}
The physical interpretation of these boundary conditions is provided
by the following Hamiltonians with external fields,
\begin{equation}
	H_\pm = -\partial_x^2 + V_0\Theta(-x) + \tilde g_\pm
	\partial_x\delta(x)\partial_x
\end{equation}
which has eigenfunctions as in Eq.\ (\ref{scatteringAnsatz2}) but with
\begin{equation}
	B_\pm = \frac{\ii k + \omega/(1+\omega \tilde g_\pm) }{\ii k
- \omega/(1+\omega \tilde g_\pm)} \quad \mbox{ and } \quad \omega =
\sqrt{V_0 - k^2}, 
\end{equation}
which converge to $A_\pm$ for $V_0\to\infty$ provided that, for example,  
\begin{equation}
	\tilde g_+ = \sqrt{V_0} \quad \mbox{ and } \quad \tilde g_- =
	2\lambda .
\end{equation}

\appende

\end{document}